\documentclass[aps, pre, reprint, floatfix, superscriptaddress]{revtex4-1}

\usepackage{amssymb, amsmath, graphicx,verbatim, color, natbib} 

\begin{document}

\title{Translational Mobilities of Proteins in Nanochannels: 
A Coarse-Grained Molecular Dynamics Study}

\author{Navaneeth Haridasan} 
\affiliation{Department of Applied Mechanics,
Indian Institute of Technology Madras, Chennai, India}

\author{Sridhar Kumar Kannam} 
\affiliation{Faculty of Science, Engineering and Technology, 
Swinburne University of Technology, Melbourne, Australia}
\affiliation{School of Sciences, RMIT University, Melbourne,
Victoria 3001, Australia.}
\author{Santosh Mogurampelly}
\affiliation{Institute for Computational
Molecular Science, Temple University, Philadelphia, United States}

\author{Sarith P. Sathian}
\altaffiliation{sarith@iitm.ac.in} 
\affiliation{Department of Applied Mechanics, Indian
Institute of Technology Madras, Chennai, India}

\date{\today}

\begin{abstract} 
We investigated the translation of a protein through
model nanopores using coarse-grained~(CG) non-equilibrium molecular
dynamics~(NEMD) simulations and compared the mobilities 
with those obtained from previous 
coarse-grained equilibrium molecular dynamics model.
We considered the effects of nanopore confinement and external force 
on the translation of streptavidin through nanopores of dimensions 
representative of experiments. 
As the nanopore radius approaches the protein hydrodynamic radius,
$r_\mathrm{h}/r_\mathrm{p}\rightarrow 1$ 
(where $r_\mathrm{h}$ is the hydrodynamic radius of protein 
and $r_\mathrm{p}$ is the pore radius), the translation
times are observed to increase by 2 orders of magnitude.
 The translation times are found to be in good 
agreement with one-dimensional biased diffusion
model. The results presented in this paper provide
 useful insights on nanopore designs intended to control the motion of 
biomolecules.
\end{abstract} 

\maketitle

\section{Introduction} 
\label{sec:introduction} 

Macromolecular
transport is ubiquitous in living organisms where biopolymers such as
proteins move from one region to another in a crowded
environment~\cite{muthukumar2011investigations,kumar2011biopolymers}. 
Initial experimental observations of poly(ethylene glycol)~\cite{
bezrukov1994counting} and single stranded DNA
~\cite{kasianowicz1996characterization} translocation in naturally occurring 
ion channels have generated huge interests in polymer translocations~\cite{
muthukumar2005simulations,wanunu2012nanopores,haque2013solid}. Since then, 
understanding the macromolecular basis of polymer translocation has
emerged as a major research activity~\cite{yeh2010single,moerner1989optical,
kannam2013nanosensors, santosh2012interaction}.

Many experimental techniques have been developed in the past few decades
for isolating and characterizing biopolymers at the single 
molecular level~\cite{walt2013optical,deniz2008single}.
Nanopore sensors based on resistive pulse sensing technique are cost-effective 
and relatively easy to use compared to the biological assays for investigating 
at the single molecules level in real time~\cite{haque2013solid,
keyser2011controlling}.
These devices measure the variation in ionic current when a biomolecule 
contained in an electrolyte solution translocates through a nanopore. The 
properties of molecules such as size, charge, conformation, concentration, etc. 
can be inferred by measuring the translocation time, drop in the current and 
frequency of translocation events~\cite{wanunu2012nanopores,kim2014@geometric,
carson2015challenges, yeh2010single, venkatesan2011nanopore,
oukhaled2012sensing, muthukumar2005simulations, ledden2011sensing}. These 
sensors have a wide range of potential applications in drug screening and 
delivery, pharmacology, molecular biology etc.~\cite{haque2013solid,
wanunu2012nanopores,kannam2013nanosensors, harrer2015label}.

Several experiments have investigated macromolecular
translocation and sensing using synthetic
nanopores~\cite{mohammad2012protein,larkin2014high,
bonome2015multistep,plesa2013fast}. Large translocation velocities of
biomolecules poses serious challenge to nanopore sensors
\cite{venkatesan2011nanopore,feng2015nanopore,
oukhaled2012sensing,carson2015challenges,maitra2012recent}.
For instance, it is infeasible to detect high velocity translocation events due
to the practical limitations such as insufficient bandwidth of the sensing 
devices~\cite{plesa2013fast}. Therefore, it is imperative to either enhance the
temporal resolution of detection or slow down the velocity, 
broadening the use of nanopore sensing  technology to a wide range of 
applications. For the latter, a fundamental understanding of the factors 
governing the directional motion of biomolecule is necessary 
and hence this study.

The first theoretical work on polymer translocation~\cite{
sung1996polymer} was concurrently published with the first experimental work on
macromolecular translocation~\cite{kasianowicz1996characterization}. This 
theoretical work was further developed shortly afterwards~\cite{
muthukumar1999polymer}. Both these models  considered the polymer translocation 
as one-dimensional diffusion problem of a long and flexible polymer chain 
assuming it to be in quasi-equilibrium.
But in typical translocation experiments, the macromolecular movement in 
nanopore is always assisted by an external force, hence it will not comply
with quasi-static assumption~\cite{kantor2004anomalous}. A general formalism 
for the forced translocation phenomena based on force balance of drag and 
applied 
force was proposed later~\cite{storm2005fast}.  
 
Researchers have modeled protein translocation by 1-D Langevin equation~\cite{
muthukumar2005simulations,
ledden2011sensing,muthukumar2014communication} and 
reported that the translocation phenomenon is mainly influenced by hydrodynamic
drag experienced by the molecule inside 
the nanopore as well as its interaction with the nanopore.  
Therefore, regulating the nanopore interactions along with increasing solvent 
viscosity, reducing the temperature, etc.~\cite{fologea2005slowing} could be 
a promising approach towards efficient sensing. We refer to 
Keyser~\cite{keyser2011controlling} for a comprehensive review on strategies 
in controlling the nanopore transport.

Kannam \textit{et al.} ~\cite{
kannam2017translational} used coarse grained equilibrum 
molecular dynamics(CGEMD) simulations to examine the effect of 
hydrodynamics on the diffusion of proteins confined in nanopores.
It was demonstrated that choosing comparable sizes of pore to the 
protein is particularly advantageous for reducing the protein diffusion and 
hence to increase the sensor efficiency.~In
this work, we apply a range of external forces on the protein and 
examine the translational motion. The external force can be regarded as an 
overall effect of electrophoretic force on the protein. The effects arising 
from partial atomic charges on different segments of proteins are not considered
~\cite{ledden2011sensing}. Comparison of the translational mobility influenced 
by external force and pore diameter with those of CGEMD model mentioned above 
and translation time distribution correspondence with a continuum model forms the 
core of this research. 

\section{Simulation Details} 
\label{sec:simulations}

The interactions between coarse-grained beads are modeled using
the Martini force field ~\cite{marrink2004coarse,marrink2007martini}.
The protein, streptavidin (PDB code : 4JO6~\cite{Barrette-Ng2013the}) 
is coarse-grained at the residue level and
each solvent bead represents four water molecules.
The protein is solvated in a solvent box with its center of mass (CM) 
tethered to the box center. The system is equilibrated in NPT ensemble
at 300~K temperature and 1~bar pressure using Berendsen 
thermostat and barostat respectively. In the next step, the nanopore is created
by freezing the solvent beads outside the required pore region in such a way 
that the pore and $z$ axis are aligned. The pore length is kept fixed at 20~nm 
and the radius $r_\mathrm{p}$ is varied from  4 to 12.5~nm to examine the 
effect of confinement on the mobility. Periodic 
boundary conditions were applied in all 3 directions making the pore infinitely
long for all the simulation cases. Considering the fact that 
protein is moving through an infinite length pore, the term translation is used
throughout this study in lieu of translocation found in nanopore literatures. 

The protein has a radius of gyration, $r_\mathrm{g}$ of 
$\sim$2.2
 nm and the maximum length of protein $m$, considering all its orientations is 
$\sim$6.5 nm. For the comparison with nanopore, protein size is 
quantified throughout the study in terms of hydrodynamic radius, $r_\mathrm{h}$.
The hydrodynamic or Stokes radius is defined as the radius of a hard sphere 
which have a bulk diffusion coefficient equivalent to the protein under similar
simulation conditions. $r_h$ is estimated as 3.22 nm from Stokes-Einstein 
relationship, $D_t^{\infty}={k_bT}/{6\pi\eta r_h}$, where the 
viscosity~\cite{kannam2017translational} $\eta$ (1.01 cP) of the solvent 
and finite-size corrected bulk 
diffusion coefficient~\cite{dunweg1993molecular} $D_t^{\infty}$ 
(0.0682 $\mathrm{nm}^2$/ns) were 
calculated from equilibrium molecular dynamics simulations of protein in bulk. 
The protein is forced through the 
pore by applying external forces corresponding to accelerations
in the range $1.0\times10^{-3}$ to $5.0\times10^{-3}$ 
$\mathrm{nm}/\mathrm{ps}^2$.
For brevity, we use the values of acceleration throughout this 
paper to represent the force applied on the protein.
To avoid the protein adsorption to the nanopore, the protein's CM 
is tethered in the radial direction to the pore axis by using a harmonic 
potential with a force constant 12~kcal~$\mathrm{mol}^{-1}$ \r{A}$^{-2}$. 
Hence the protein is free to move in the $z$ direction 
and restrained to move in $x$ and $y$ directions. No restraints are applied 
on the rotational degree of freedom of the protein. The production 
simulations are carried out in NVT ensemble with a time step of 20~fs for 
200~ns using GROMACS~\cite{berendsen1995gromacs} simulation package. The 
simulation models and protocols are adapted from 
Kannam~\textit{et al}~\cite{kannam2017translational}. 

\begin{figure}[!h] 
\includegraphics[width=0.4\textwidth]{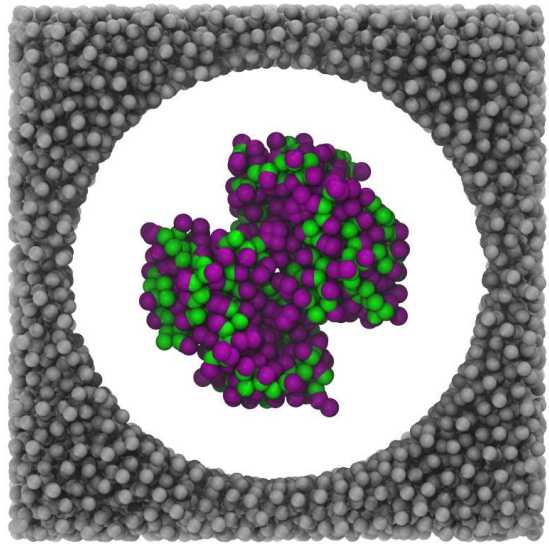}
\caption{Coarse-grained model of the protein, streptavidin (PDB
code : 4JO6~\cite{Barrette-Ng2013the}), 
confined in a 
nanopore of radius 5~nm. The nanopore, protein backbone and side chains are 
colored in grey, green, and purple respectively. The solvent is not shown for
visual clarity.} 
\label{fig:schematic} 
\end{figure}

\section{Results and Discussion} 
\label{sec:results} 

\subsection{Conformational stability of the protein} 

In the translocation experiments at physiological conditions, 
globular proteins are generally proclaimed to maintain its 
conformation without any significant changes~\cite{fologea2007electrical,
ledden2011sensing}. 
But in NEMD simulations, the proteins are reported to undergo conformational 
changes due to the large external forces~\cite{kannam2014sensing,
talaga2009single,aksimentiev2009deciphering}. 
To measure the conformational stability of the protein, we measured the 
root mean square deviation (RMSD) and the radius of gyration ($r_\mathrm{g}$)
during the translation in our NEMD simulations.  

\begin{figure}[!h] 
\includegraphics[width=0.4\textwidth]{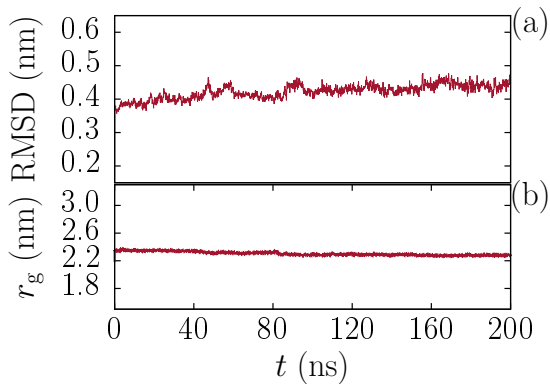}
\caption{(a) RMSD and the (b) radius of gyration ($r_\mathrm{g}$) of 
streptavidin with time during the translation at an acceleration 
of $1.0\times10^{-3}$ $\mathrm{nm}/\mathrm{ps}^2$ in NEMD simulation.} 
\label{fig:rmsdrgdistb} 
\end{figure}

RMSD is a measure of protein's conformational change which is calculated as 
the root mean square deviation of the protein's atom positions with respect 
to a reference structure.  
\begin{equation}
\mathrm{RMSD}(t)=\sqrt{\frac{1}{N}\sum_{i=1}^N({\mathbf{r}_i(t)-\mathbf{r}
_i^{\mathrm{ref}}})^2}
\end{equation} 
where $N$ is the number of atoms in the protein.  
The $\mathbf{r}_i(t)$ represents the position vector of $i^{\text{th}}$ 
atom at time $t$ and $\mathbf{r}_i^{\text{ref}}$ denotes its 
reference structure position vector. The radius of gyration, 
$r_\mathrm{g}=\sqrt{\frac{1}{N}\sum_{i=1}^N({\mathbf{r}_i-\mathbf{r}_{
\mathrm{CM}}})^2}$
of the protein on the other hand, quantifies its globular size  
providing an alternate quantification of the protein's conformational changes. 
The time evolution of RMSD and $r_\mathrm{g}$ is  shown in 
Figures~\ref{fig:rmsdrgdistb}(a) and (b) as the protein translocates through 
the pore, at an applied force corresponding to the acceleration 
$1.0\times10^{-3}$ $\mathrm{nm}/\mathrm{ps}^2$. 
We find no significant deviation in the conformational state of the protein,
indicating that the magnitudes of external forces considered in the NEMD 
simulations are reasonably appropriate
to investigate the translation phenomena. Similar observations were made at 
the other accelerations encountered in this study.  

\subsection{Translation Process and Mobilities} 

Previous studies reported that proteins could be adsorbed on the nanopore 
surface which significantly increases the translocation time, induce 
conformational changes and even trigger the  activation of
binding sites of the protein~\cite{eleonora2016role, eleonora2017proton,
rodriguez1991surface, durand2008label,carr2011modelling}. 
While it is important to understand such effects, we simplified our model
by tethering the CM of protein to pore axis with a harmonic potential, 
to primarily study the 
translation mechanism without the 
protein adsorbing
to the pore surface.  

\begin{figure}[htb]
\centering 
\includegraphics[width=0.4\textwidth]{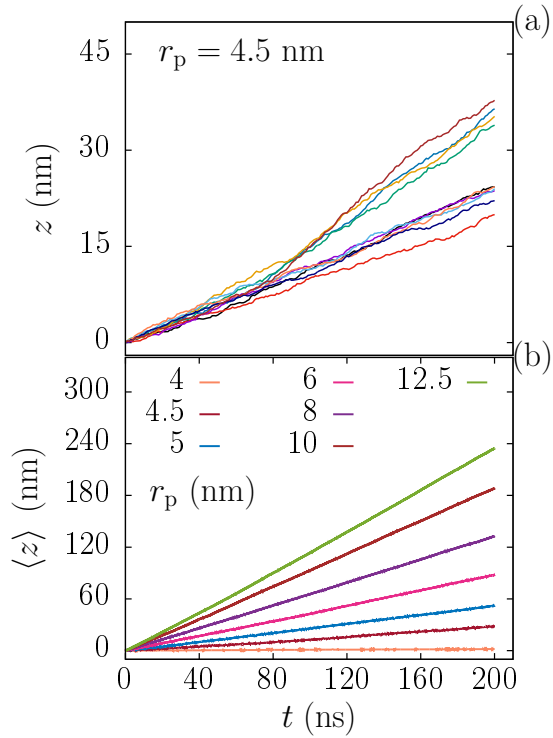}
\caption{(a) The instantaneous center of mass position of streptavidin along 
the pore axis,~$z$, with time for 10 replica simulations for a 
pore radius of 4.5~nm and at an acceleration of~$1.0\times10^{-3}$ 
$\mathrm{nm}/\mathrm{ps}^2$. (b) The
evolution of average center of mass position of streptavidin along the pore 
axis,~$\left\langle z 
\right\rangle$, at different pore radii 
ranging between 4 to 12.5~nm.}
\label{fig:position} 
\end{figure}

\begin{figure}[htb]
\centering 
\includegraphics[width=0.4\textwidth]{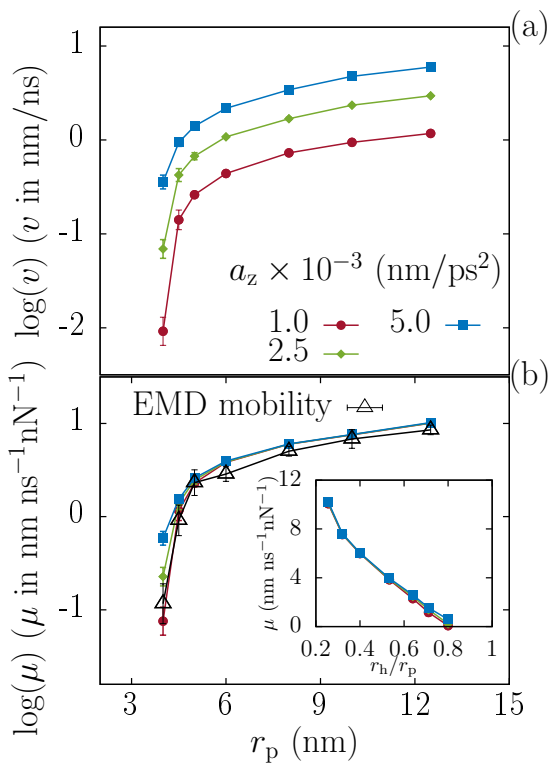}
\caption{(a) The average translation velocity $v$ of protein as a function of
the pore radius at different accelerations corresponding to 
external forces. (b) Comparison of average 
mobilities calculated from non-equilibrium MD simulations (legends are same as 
Figure 4(a)) with the Einstein-Smoluchowski predictions as in 
Equation~\ref{eq:ES} (triangles).}
\label{fig:velocity} 
\end{figure}

In Figure~\ref{fig:position}(a), we show the instantaneous CM position of 
protein along the pore axis, $z$, in 10 independent simulations for 
a pore radius of 4.5~nm and at an external force corresponding 
to acceleration of $1.0\times10^{-3}$~$\mathrm{nm}/\mathrm{ps}^2$. For a given
trajectory, we observe that $z$ exhibits local fluctuations on short 
time scales corresponding to protein's typical random walk in a thermal 
environment. Due to the external force, the protein 
translocates along the pore axis with a net velocity.
From the independent simulations for particular
values of applied external force and pore radii, the velocities based on 
$z$ were calculated and averaged to get a mean velocity. The error bars in
Figure~\ref{fig:velocity}(a) represents the standard deviation of the mean 
velocity values calculated from independent simulations. 
A t-test~\cite{student1908probable} was adopted to 
ensure that the calculated mean value represents population mean with 
sufficient confidence. For the case with largest standard deviation, it was
found that margin of error is around 10\% of the mean value at a confidence 
level of 90\%. Thus we conclude that the calculated mean value represents 
the population mean value with the error of 10\% or less at 90\% confidence.  

While the observation of protein translation is quite natural because of the 
applied force, the effects of 
pore radius on protein dynamics
are found to be significant, as shown in Figure~\ref{fig:position}(b). The 
linearity of average CM position of protein along the pore axis,~$\left\langle
z \right\rangle$ indicates that the velocity is constant 
throughout the nanopore of any radius, $r_\mathrm{p}$. The higher strength of 
protein-nanopore interactions with decreasing pore radius is seen to 
reduce the protein displacements significantly.
Notably, in a nanopore of the smaller radius considered in our simulations, the
protein is observed to displace barely about its size during the entire 200~ns
trajectory. On the other hand, for the largest radius considered, the protein
displaces approximately 100~$r_\mathrm{g}$ in 200~ns, which is
two orders greater than the average displacement 
of protein in bulk due to diffusion (based on Peclet number calculation defined 
later).

The above results demonstrate that there are considerable variations in protein 
displacements depending on the pore radius at a given applied force. To 
understand the dynamics of the protein, the mean velocity $v$ calculated using 
$\left\langle z \right\rangle$ is displayed in Figure~\ref{fig:velocity}(a) at 
different applied forces. Consistent with the variations in
$\left\langle z \right\rangle$, the velocities are seen to change 
significantly with the pore radius. The change of velocity is 
observed to span on order of magnitude below a critical pore radius, 
$r_\mathrm{p}^c$ and approaches a constant value for 
$r_\mathrm{p} > r_\mathrm{p}^c$. 
This reveals the effect of pore friction on the translation 
velocity and consequently the translation time. This is similar to the 
translocation 
dynamics of polymer chains where the pore friction increases with decrease of 
pore diameter and dominates as a finite size effect especially for a short 
polymer~\cite{ikonen2013influence}. The proportional relation between the 
translation velocity and external forces obtained from the simulations
suggests that the forces used are in the linear regime. 

Another interesting feature observed from
Figure~\ref{fig:velocity}(a) is that the qualitative differences in velocity with 
$r_\mathrm{p}$ are almost identical at given external forces. To quantify such a 
behavior, we calculated the mobility, $\mu$ as the ratio of velocity to the 
applied force and the results are displayed in Figure~\ref{fig:velocity}(b).
Following the qualitative features observed for the velocities, we find 
that the mobilities are independent of the applied force for $r_\mathrm{p}$
$>$ 4 nm, which is consistent with the experimental observation that the 
biopolymer electrophoretic mobilities are independent of the applied electric 
fields both in free solution~\cite{stellwagen2001orientation} and in a 
translocation setup~\cite{peng2004probing}. However, the mobility 
significantly decreases as $r_\mathrm{p}$ equals 4 nm or in 
general as it approaches the dimensions of 
the protein.

The translation mobilities calculated from the velocities can be directly
compared~\cite{brown1988comparison,pryamitsyn2016noncontinuum} with the 
predictions of Einstein-Smoluchowski relation:
\begin{equation}
\mu(r_\mathrm{p})={D(r_\mathrm{p})}/{k_\mathrm{B}T} 
\label{eq:ES}
\end{equation}
where $D(r_\mathrm{p})$ is the diffusion coefficient of the
protein in a nanopore of radius $r_\mathrm{p}$, $k_\mathrm{B}$ is the Boltzmann 
constant and $T$ is the temperature in Kelvin. Assuming that the
$D(r_\mathrm{p})$ is independent of the applied
force~\cite{nkodo2001diffusion}, we evaluated $\mu(r_\mathrm{p})$ by using the
respective $D(r_\mathrm{p})$ obtained under the equilibrium 
conditions~\cite{kannam2017translational}.
Interestingly, we observe an excellent agreement between the $\mu$ obtained 
from the NEMD simulations and the EMD based 
Einstein-Smoluchowski predictions.  

\begin{figure}[!h]
\includegraphics[width=0.4\textwidth]{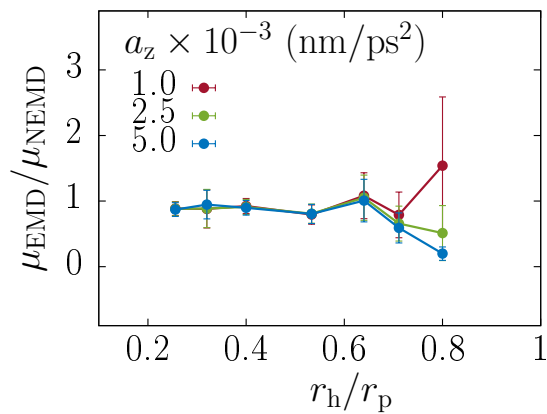}
\caption{Ratio of EMD to NEMD mobilities for different
pore sizes and applied forces. There is marked variation of the ratio with
respect to the applied forces when the pore size is closer to protein
dimensions.}
\label{fig:muratio}
\end{figure}

The comparison of mobilities from EMD and NEMD simulations in 
Figure~\ref{fig:velocity}(b) indicates that quantitatively the system exhibits 
similar behaviour in terms of diffusion and external forcing velocity when 
the pore diameter is varied. From the mean values and error bars from Figure~
\ref{fig:muratio}, it can 
be observed that EMD and NEMD mobilities lies very close to each other for 
$r_\mathrm{p}$ values other than 4 nm ($r_\mathrm{h}/r_\mathrm{p} = 0.805$). 
This equivalence is 
applicable even for large forces considered in the simulation where diffusion 
time scales are almost two orders lesser than forced translation time scales. 
(The two order difference between transport rate is quantified by Peclet 
number $Pe$ for the protein, defined as the 
ratio of forced translation rate to diffusion transport rate,
$Pe = \dfrac{r_\mathrm{h}v(r_\mathrm{p})}{D(r_\mathrm{p})}$ where  
$v(r_\mathrm{p})$ is the velocity due to external force). 
Apparently, it can be safely assumed that the equivalence between EMD and 
NEMD mobilities prevails for any external forces between equilibrium and
the forces considered in the simulation. This is not the case with pore sizes
close to $r_\mathrm{h}$ considered in the simulation as shown in Figure~
\ref{fig:velocity}(b) and \ref{fig:muratio}. There is marked variation of 
mobility  experienced by the protein with respect to the applied force for 
$r_\mathrm{p}$ value 4 ($r_\mathrm{h}/r_\mathrm{p} = 0.805$). This indicates 
that the mobility values depend on the force 
applied, which signifies the presence of excessive drag compared to other pores
in the study. Similar observation has been made before in the literature for 
DNA translocation~\cite{carson2014smooth}.

\subsection{Probability distribution function of translation 
times} 

Numerous attempts were made previously to theoretically formulate the 
distribution of translocation times~\cite{
lubensky1999driven,berezhkovskii2003translocation,talaga2009single,ling2013on}.
Among these, Talaga et al. ~\cite{talaga2009single} proposed a biased 1-D
Fokker-Planck diffusion model with 
relatively realistic boundary conditions which was later corrected as 
Schr\"odinger's first passage probability distribution function 
(FP-PDF)~\cite{ling2013on,talaga2013correction}. 

Following the arguments and the treatment of Talaga and Li
~\cite{talaga2009single,ling2013on,talaga2013correction}, we use 1-D
FP-PDF to broadly describe the translation of protein 
through nanopore under the influence of a constant force. Since 
the mobilities from EMD and NEMD are equivalent from Figure~\ref{fig:velocity}
(b), we use Equation~\ref{eq:ES} and velocity-force relation of mobility 
from NEMD to replace diffusion coefficient $D$ of the original 1-D FP-PDF.   
For the translation, the protein has to travel a distance $l$ in time $t$, 
with an induced velocity $v$. Due to the 
stochastic nature of translation process, the time required to translocate 
distance $l$ assumes a probability distribution, $p(t)$ given by : 

\begin{equation} 
p(t)=\bigg[\frac{l^2 F}{4\pi{k_\mathrm{B}T}vt^3}\bigg]^{1/2} e^{{-}\frac{F(l-tv)^2}{4{k_\mathrm{B}T}vt}}
\label{eq:poft}
\end{equation}
In the above Equation, we chose $l$ = 50 nm and $F$ as the 
external force corresponding to the acceleration of~$1.0\times10^{-3}$
$\mathrm{nm}/\mathrm{ps}^2$. The velocity is obtained by scaling the EMD
mobility in Equation~\ref{eq:ES} with $F$.  

\begin{figure}[!h] 
\includegraphics[width=0.4\textwidth]{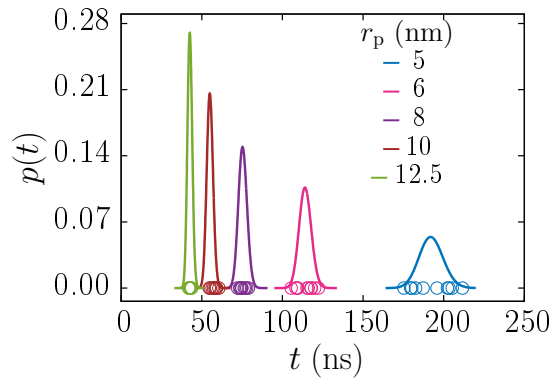}
\caption{Probability distribution of translation times at different 
$r_\mathrm{p}$ obtained numerically using Equation~\ref{eq:poft} for 
streptavidin in nanopores for $F$ correspeonding to external 
acceleration of ~$1.0\times10^{-3}$ $\mathrm{nm}/\mathrm{ps}^2$ . The circles 
indicate the translation times 
calculated with the velocities obtained from 10 independent NEMD trajectories, and agrees well with the EMD
based predictions using Equation \ref{eq:poft}. Note that due
to the limitations on computational requirements (a typical problem associated
with simulations), we were not able to compute $p(t)$ directly from NEMD
simulations which require roughly $10^3-10^4$ independent trajectories.}
\label{fig:fppdf} 
\end{figure} 

Figure~\ref{fig:fppdf} depicts $p(t)$ obtained numerically for streptavidin at
different values of $r_\mathrm{p}$. We observe that while the shape of $p(t)$ 
remains unaffected, the width and peak height changes significantly with 
$r_\mathrm{p}$. The range of translation times calculated directly from 
NEMD velocities (circles) lies within the width estimated using 
Equation~\ref{eq:poft}. Moreover, the NEMD values are centered around the 
higher probability region for most cases of $r_\mathrm{p}$. 
However, as the protein is restrained to the nanopore axis
the translation process is smooth without the adsorption on pore walls. 
 Also the capture phenomena of the protein 
is not modeled within our framework, hence unsuccessful translocation
events or collisions~\cite{wanunu2008dna} are not accounted during entire 
simulation.
This may lead to an under-representation of the long tail of translation 
times observed in experiments~\cite{kasianowicz1996characterization,
fologea2007electrical,plesa2013fast}.

\begin{figure}[!h]
\includegraphics[width=0.4\textwidth]{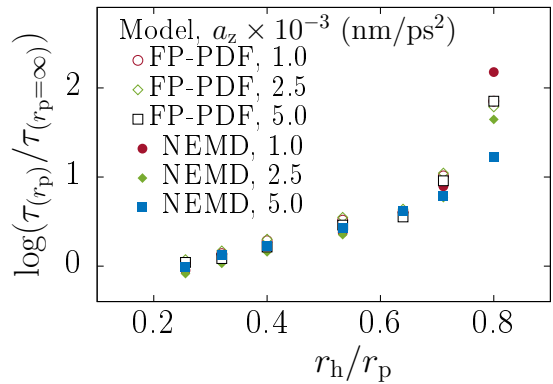}
\caption{Translation time dependency on
nanopore radius. Data for NEMD simulations and 1-D diffusion
model at different applied forces collapses on a universal curve.
Translation times evaluated using Equation~\ref{eq:tau} (1-D Model)
agrees with those obtained from the NEMD simulations.}
\label{fig:translationtime}
\end{figure}

Another interesting property relevant for the translation experiments is the
mean translation time, $\tau$ which can be calculated numerically by using 
the probability distributions as: 
\begin{equation}
\tau=\frac{\int_{0}^{\infty} t p(t)dt}{\int_{0}^{\infty} p(t) dt}
\label{eq:tau} 
\end{equation}
A comparison of $\tau$ calculated using Equation~\ref{eq:tau} and those 
obtained directly from NEMD simulations is presented in 
Figure~\ref{fig:translationtime} at different values of 
$r_\mathrm{p}$ and external forces.
In the Figure, $\tau_{(r_\mathrm{p} = \infty)}$ denotes the time required for 
protein to travel $l$ for $r_\mathrm{p} = \infty$, which is extracted by fitting
the data to 
$\tau_0 e^{\lambda(r_\mathrm{h}/r_\mathrm{p})}+\tau_{(r_\mathrm{p} = \infty)}$, 
where $\tau_0$ and 
$\lambda$ are fitting parameters. The mean translation times  
calculated from NEMD simulations are seen to decrease rapidly with nanopore 
radius and approaches $\tau_{(r_\mathrm{p} = \infty)}$ as shown in 
Figure~\ref{fig:translationtime}. The translation times predicted from the 
1-D FP-PDF model agrees well with the NEMD results for most values of 
$r_\mathrm{p}$, though it differs for pores similar in size of the protein. 
Moreover, the changes in $\tau$ with $r_\mathrm{p}$ are consistent with the 
results presented in Figure~\ref{fig:velocity}(a). Additionally, the qualitative
features of $\tau$ are very similar to those reported for voltage-driven DNA 
translocation through solid-state nanopores~\cite{wanunu2008dna}. 

The ability of 1-D diffusion model in predicting the qualitative features of
$\tau$ and capturing the physics of translation are promising despite ignoring
details such as electroosmotic gradients, electrostatic charge
distribution,the shape of protein, etc. Overall, the qualitative agreement
between the 1-D diffusion model and NEMD simulations helps in rationalizing the
significance of diffusion dependent mechanisms in governing the translation of
proteins in nanopores.

\subsection{Nature of Drag on the Protein}

An interesting
question will be the source of increase in drag
on the protein as the pore diameter approaches hydrodynamic radius,
$r_\mathrm{h}$. From   
Figure~\ref{fig:translationtime} and Figure~\ref{fig:muratio},
 it is evident that for the pores with least two diameters considered in the 
simulation, there is a disparity with respect to velocity or
time required for the protein translation. This is inline with the
diffusion coefficient variation related to change in pore size expressed
in previous EMD study\cite{kannam2017translational}. While most of the protein 
translation times corresponding 
to different pore sizes follow the frictional drag relationship 
$(\sim  r_\mathrm{h}/(r_\mathrm{p}-r_\mathrm{h}))$~\cite{wanunu2008dna}, 
cases of $r_\mathrm{p}$ with
4 and 4.5 ($r_\mathrm{h}/r_\mathrm{p}$ value 0.805 and 0.716 respectively) 
show considerable deviation from this relation. Thus, it can be
inferred that, for large pores the drag is mainly a hydrodynamic effect
but as $ r_\mathrm{h}$ approaches $r_\mathrm{p}$, there may be additional
effects which contribute to the drag on the protein.
\begin{figure}[!h]
\includegraphics[width=0.4\textwidth]{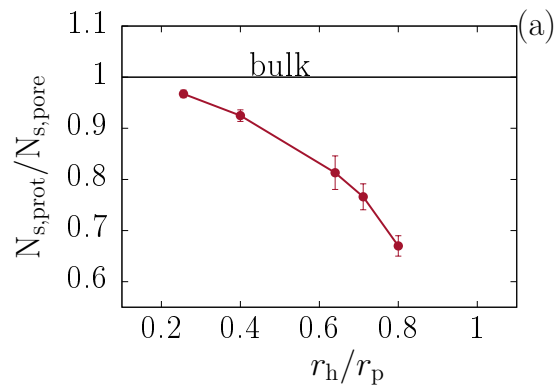}
\caption{ Ratio of number of solvent beads per nm for the 
distance occupied by the protein in $z$ direction $\mathrm{N}_{\mathrm{s,prot}}$,
to the number of solvent beads per nm for rest of the pore length 
$\mathrm{N}_{\mathrm{s,pore}}$ versus pore radii. The ratio indirectly
indicates the volume occupied by the protein in comparison to solvent,
if pore length is exactly $z$ dimension of protein.}
\label{fig:nsprot}
\end{figure}

The ratio of number of solvent beads per nm for the
distance occupied by the protein in $z$ direction $\mathrm{N}_{\mathrm{s,prot}}$
to the number of solvent beads per nm for rest of the pore length
$\mathrm{N}_{\mathrm{s,pore}}$ is plotted against different pore radii 
in Figure~\ref{fig:nsprot}. This indicates the number of solvent
beads occupying the pore volume along with the protein, if the pore length is 
exactly the $z$ dimension
of protein. From Figure~\ref{fig:nsprot}, it is observed that 
for a given $z$ dimension of the protein, number of solvent beads are just 
two times the number of protein beads for 
the least pore diameter. This leads to the inference that the protein occupies 
considerable volume inside the pore in comparison to solvent beads. Most 
importantly, since the maximum dimension $m$ of the protein is around 6.5 nm 
throughout the simulation, for the least pore diameter case, the solvent may 
form a thin film between pore and protein or sparsely distributed solvent 
clusters may exist between protein and nanopore surface. 

\begin{figure}[!h]
\includegraphics[width=0.4\textwidth]{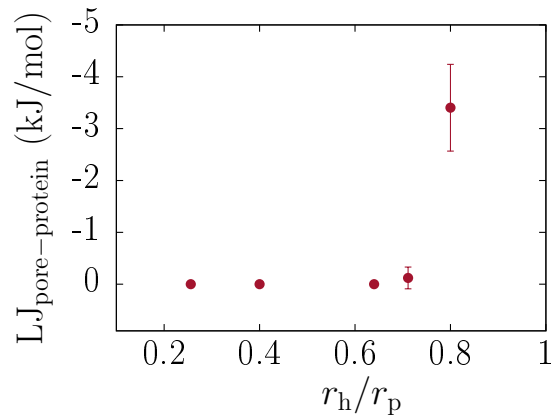}
\caption{ Non bonded interaction between pore and protein
for different pore size. $r_\mathrm{p}$ 4 and 4.5 nm ($r_\mathrm{h}/r_\mathrm{p}$ value 0.805 and 0.716 respectively) exhibits non negligible
values of interaction compared to other values of $r_\mathrm{p}$ considered. }
\label{fig:lj}
\end{figure}

Apart from above analysis, accounting the non-bonded interaction
between pore and protein as shown in Figure~\ref{fig:lj} reveals that there is a
non negligible interaction between pore and protein for the case of two 
nanopores with the least diameters considered in the study. Thus the drag may 
be caused due to the presence of a thin film of solvent or due to direct 
non-bonded interaction between pore and protein or both.
While it can be speculated from above analyses that the drag increase is mainly due to the non-bonded interaction between protein and pore, further 
analyses are required for a conclusive understanding of the underlying 
mechanisms.  

\section{Concluding Remarks and Outlook}
 
To conclude, the mechanisms governing
protein translation through nanopores are investigated in this work by
considering a coarse-grained protein (streptavidin) with model
nanopores using molecular dynamics simulations. To accelerate the translation
process within computationally tractable timescales, we applied external forces 
of different magnitudes on proteins. Nanopores of varying radii ranging from 4
to 12.5~nm ($r_\mathrm{h}/r_\mathrm{p}$ value 0.805 to 0.258 
respectively) that are comparable to the experimental length scales are considered 
to understand the effects of pore size. We simulated the translation process 
for 10 different initial configurations to generate trajectories of 200~ns in 
each case.

One of the important outcomes of this paper is that  
the mobilities 
calculated from the NEMD simulations are comparable to those computed from  
EMD simulations indicating the 
similarity in system response to a wide range 
of forces starting from values close to equilibrium. 
Also, the diffusion coefficient or the mobilities depends strongly
on the pore size which can be used to control the 
directional or diffusional motion of the protein in nanopores. As a 
consequence, the mean translation times are observed to be increasing rapidly
with decreasing pore sizes which is in good agreement with previous experimental
and simulation results~\cite{wanunu2008dna,menais2016polymer,menais2017polymer}. 

A stochastic model based on 1-D 
biased diffusion equation was used to study the translation of proteins in 
conjunction with MD simulations. The results of
such a simplistic model are seen to corroborate the large scale coarse-grained
simulations and provide further significant insights. Specifically, the 
probability distributions of the translation time indicated larger standard 
deviations in translation time for smaller pores which are consistent with 
typical experimental reports~\cite{wanunu2008dna}. More interestingly, the 
average translation times predicted using the stochastic model decreases 
rapidly with the pore size, is in excellent agreement with NEMD simulations.
Also a two order increase of translational time signifies 
presence of additional effects contributing to drag experienced by the protein,
the nature of which needs to be revealed through a separate detailed study. 
The results presented in this paper provide useful insights on the nanopore 
design for several applications. For instance, the nanopore may be specifically
designed to control protein diffusivity for accurate sequencing or similar 
applications.

Recent atomistic simulations have reported that the changes in water density
fluctuations in a proton channel can potentially initiate active binding sites
for the target molecules subjected to translation
\cite{eleonora2016role,eleonora2017proton}. In the drug design community, such
an observation is expected to generate significant interest in the quest for 
potential inhibitors of the proton channel. Similar phenomena could arise when 
protein is subjected to translation in thin nanopores, where the degree of 
hydration of protein becomes crucial. Explicitly, the preferential interactions
between nanopore and protein's residues play a major role in the translation
process. We note that while the protein is restrained along the nanopore axis in
the present model, it is highly desirous to understand the importance of solvent
mediated dry-wet mechanism of protein translation process and the manner in 
which they are influenced by the nanopore size.



\begin{thebibliography}{62}%
\makeatletter
\providecommand \@ifxundefined [1]{%
 \@ifx{#1\undefined}
}%
\providecommand \@ifnum [1]{%
 \ifnum #1\expandafter \@firstoftwo
 \else \expandafter \@secondoftwo
 \fi
}%
\providecommand \@ifx [1]{%
 \ifx #1\expandafter \@firstoftwo
 \else \expandafter \@secondoftwo
 \fi
}%
\providecommand \natexlab [1]{#1}%
\providecommand \enquote  [1]{``#1''}%
\providecommand \bibnamefont  [1]{#1}%
\providecommand \bibfnamefont [1]{#1}%
\providecommand \citenamefont [1]{#1}%
\providecommand \href@noop [0]{\@secondoftwo}%
\providecommand \href [0]{\begingroup \@sanitize@url \@href}%
\providecommand \@href[1]{\@@startlink{#1}\@@href}%
\providecommand \@@href[1]{\endgroup#1\@@endlink}%
\providecommand \@sanitize@url [0]{\catcode `\\12\catcode `\$12\catcode
  `\&12\catcode `\#12\catcode `\^12\catcode `\_12\catcode `\%12\relax}%
\providecommand \@@startlink[1]{}%
\providecommand \@@endlink[0]{}%
\providecommand \url  [0]{\begingroup\@sanitize@url \@url }%
\providecommand \@url [1]{\endgroup\@href {#1}{\urlprefix }}%
\providecommand \urlprefix  [0]{URL }%
\providecommand \Eprint [0]{\href }%
\providecommand \doibase [0]{http://dx.doi.org/}%
\providecommand \selectlanguage [0]{\@gobble}%
\providecommand \bibinfo  [0]{\@secondoftwo}%
\providecommand \bibfield  [0]{\@secondoftwo}%
\providecommand \translation [1]{[#1]}%
\providecommand \BibitemOpen [0]{}%
\providecommand \bibitemStop [0]{}%
\providecommand \bibitemNoStop [0]{.\EOS\space}%
\providecommand \EOS [0]{\spacefactor3000\relax}%
\providecommand \BibitemShut  [1]{\csname bibitem#1\endcsname}%
\let\auto@bib@innerbib\@empty
\bibitem [{\citenamefont {Muthukumar}(2011)}]{muthukumar2011investigations}%
  \BibitemOpen
  \bibfield  {author} {\bibinfo {author} {\bibfnamefont {M.}~\bibnamefont
  {Muthukumar}},\ }\enquote {\bibinfo {title} {Polymer translocation},}\ in\
  \href@noop {} {\emph {\bibinfo {booktitle} {Investigations on the theory of
  Brownian movement}}}\ (\bibinfo  {publisher} {CRC Press},\ \bibinfo {year}
  {2011})\BibitemShut {NoStop}%
\bibitem [{\citenamefont {Kumar}\ \emph {et~al.}(2011)\citenamefont {Kumar},
  \citenamefont {Lansac}, \citenamefont {Glaser},\ and\ \citenamefont
  {Maiti}}]{kumar2011biopolymers}%
  \BibitemOpen
  \bibfield  {author} {\bibinfo {author} {\bibfnamefont {H.}~\bibnamefont
  {Kumar}}, \bibinfo {author} {\bibfnamefont {Y.}~\bibnamefont {Lansac}},
  \bibinfo {author} {\bibfnamefont {M.~A.}\ \bibnamefont {Glaser}}, \ and\
  \bibinfo {author} {\bibfnamefont {P.~K.}\ \bibnamefont {Maiti}},\ }\href@noop
  {} {\bibfield  {journal} {\bibinfo  {journal} {Soft Matter}\ }\textbf
  {\bibinfo {volume} {7}},\ \bibinfo {pages} {5898} (\bibinfo {year}
  {2011})}\BibitemShut {NoStop}%
\bibitem [{\citenamefont {Bezrukov}\ \emph {et~al.}(1994)\citenamefont
  {Bezrukov}, \citenamefont {Vodyanoy},\ and\ \citenamefont
  {Parsegian}}]{bezrukov1994counting}%
  \BibitemOpen
  \bibfield  {author} {\bibinfo {author} {\bibfnamefont {S.~M.}\ \bibnamefont
  {Bezrukov}}, \bibinfo {author} {\bibfnamefont {I.}~\bibnamefont {Vodyanoy}},
  \ and\ \bibinfo {author} {\bibfnamefont {V.~A.}\ \bibnamefont {Parsegian}},\
  }\href@noop {} {\bibfield  {journal} {\bibinfo  {journal} {Nature}\ }\textbf
  {\bibinfo {volume} {370}},\ \bibinfo {pages} {279} (\bibinfo {year}
  {1994})}\BibitemShut {NoStop}%
\bibitem [{\citenamefont {Kasianowicz}\ \emph {et~al.}(1996)\citenamefont
  {Kasianowicz}, \citenamefont {Brandin}, \citenamefont {Branton},\ and\
  \citenamefont {Deamer}}]{kasianowicz1996characterization}%
  \BibitemOpen
  \bibfield  {author} {\bibinfo {author} {\bibfnamefont {J.}~\bibnamefont
  {Kasianowicz}}, \bibinfo {author} {\bibfnamefont {E.}~\bibnamefont
  {Brandin}}, \bibinfo {author} {\bibfnamefont {D.}~\bibnamefont {Branton}}, \
  and\ \bibinfo {author} {\bibfnamefont {D.}~\bibnamefont {Deamer}},\
  }\href@noop {} {\bibfield  {journal} {\bibinfo  {journal} {Proc. Natl. Acad.
  Sci. USA}\ }\textbf {\bibinfo {volume} {93}},\ \bibinfo {pages} {13770}
  (\bibinfo {year} {1996})}\BibitemShut {NoStop}%
\bibitem [{\citenamefont {Kong}\ and\ \citenamefont
  {Muthukumar}(2005)}]{muthukumar2005simulations}%
  \BibitemOpen
  \bibfield  {author} {\bibinfo {author} {\bibfnamefont {C.~Y.}\ \bibnamefont
  {Kong}}\ and\ \bibinfo {author} {\bibfnamefont {M.}~\bibnamefont
  {Muthukumar}},\ }\href@noop {} {\bibfield  {journal} {\bibinfo  {journal} {J.
  Am. Chem. Soc.}\ }\textbf {\bibinfo {volume} {127}},\ \bibinfo {pages}
  {18252} (\bibinfo {year} {2005})}\BibitemShut {NoStop}%
\bibitem [{\citenamefont {Wanunu}(2012)}]{wanunu2012nanopores}%
  \BibitemOpen
  \bibfield  {author} {\bibinfo {author} {\bibfnamefont {M.}~\bibnamefont
  {Wanunu}},\ }\href@noop {} {\bibfield  {journal} {\bibinfo  {journal} {Phys.
  Life Rev.}\ }\textbf {\bibinfo {volume} {9}},\ \bibinfo {pages} {125}
  (\bibinfo {year} {2012})}\BibitemShut {NoStop}%
\bibitem [{\citenamefont {Haque}\ \emph {et~al.}(2013)\citenamefont {Haque},
  \citenamefont {Li}, \citenamefont {Wu}, \citenamefont {Liang},\ and\
  \citenamefont {Guo}}]{haque2013solid}%
  \BibitemOpen
  \bibfield  {author} {\bibinfo {author} {\bibfnamefont {F.}~\bibnamefont
  {Haque}}, \bibinfo {author} {\bibfnamefont {J.}~\bibnamefont {Li}}, \bibinfo
  {author} {\bibfnamefont {H.-C.}\ \bibnamefont {Wu}}, \bibinfo {author}
  {\bibfnamefont {X.-J.}\ \bibnamefont {Liang}}, \ and\ \bibinfo {author}
  {\bibfnamefont {P.}~\bibnamefont {Guo}},\ }\href@noop {} {\bibfield
  {journal} {\bibinfo  {journal} {Nano Today}\ }\textbf {\bibinfo {volume}
  {8}},\ \bibinfo {pages} {56} (\bibinfo {year} {2013})}\BibitemShut {NoStop}%
\bibitem [{\citenamefont {Yeh}\ \emph {et~al.}(2005)\citenamefont {Yeh},
  \citenamefont {Chao}, \citenamefont {Ho},\ and\ \citenamefont
  {Wang}}]{yeh2010single}%
  \BibitemOpen
  \bibfield  {author} {\bibinfo {author} {\bibfnamefont {H.-C.}\ \bibnamefont
  {Yeh}}, \bibinfo {author} {\bibfnamefont {S.-Y.}\ \bibnamefont {Chao}},
  \bibinfo {author} {\bibfnamefont {Y.-P.}\ \bibnamefont {Ho}}, \ and\ \bibinfo
  {author} {\bibfnamefont {T.-H.}\ \bibnamefont {Wang}},\ }\href@noop {}
  {\bibfield  {journal} {\bibinfo  {journal} {Curr. Pharm. Biotech.}\ }\textbf
  {\bibinfo {volume} {6}},\ \bibinfo {pages} {453} (\bibinfo {year}
  {2005})}\BibitemShut {NoStop}%
\bibitem [{\citenamefont {Moerner}\ and\ \citenamefont
  {Kador}(1989)}]{moerner1989optical}%
  \BibitemOpen
  \bibfield  {author} {\bibinfo {author} {\bibfnamefont {W.~E.}\ \bibnamefont
  {Moerner}}\ and\ \bibinfo {author} {\bibfnamefont {L.}~\bibnamefont
  {Kador}},\ }\href@noop {} {\bibfield  {journal} {\bibinfo  {journal} {Phys.
  Rev. Lett.}\ }\textbf {\bibinfo {volume} {62}},\ \bibinfo {pages} {2535}
  (\bibinfo {year} {1989})}\BibitemShut {NoStop}%
\bibitem [{\citenamefont {Kannam}\ \emph {et~al.}(2013)\citenamefont {Kannam},
  \citenamefont {Downton}, \citenamefont {Gunn}, \citenamefont {Kim},
  \citenamefont {Rogers}, \citenamefont {Schieber}, \citenamefont {Baldauf},
  \citenamefont {Wagner}, \citenamefont {Daniel}, \citenamefont {Bathgate},\
  and\ \citenamefont {Harrer}}]{kannam2013nanosensors}%
  \BibitemOpen
  \bibfield  {author} {\bibinfo {author} {\bibfnamefont {S.~K.}\ \bibnamefont
  {Kannam}}, \bibinfo {author} {\bibfnamefont {M.~T.}\ \bibnamefont {Downton}},
  \bibinfo {author} {\bibfnamefont {N.}~\bibnamefont {Gunn}}, \bibinfo {author}
  {\bibfnamefont {S.~C.}\ \bibnamefont {Kim}}, \bibinfo {author} {\bibfnamefont
  {P.~R.}\ \bibnamefont {Rogers}}, \bibinfo {author} {\bibfnamefont
  {C.}~\bibnamefont {Schieber}}, \bibinfo {author} {\bibfnamefont {J.~S.}\
  \bibnamefont {Baldauf}}, \bibinfo {author} {\bibfnamefont {J.~M.}\
  \bibnamefont {Wagner}}, \bibinfo {author} {\bibfnamefont {S.}~\bibnamefont
  {Daniel}}, \bibinfo {author} {\bibfnamefont {R.}~\bibnamefont {Bathgate}}, \
  and\ \bibinfo {author} {\bibfnamefont {S.}~\bibnamefont {Harrer}},\ }in\
  \href@noop {} {\emph {\bibinfo {booktitle} {SPIE Micro Nano Materials,
  Devices, and Systems}}}\ (\bibinfo {organization} {Int. Soc. Opt. Phot.},\
  \bibinfo {year} {2013})\ p.\ \bibinfo {pages} {89230I}\BibitemShut {NoStop}%
\bibitem [{\citenamefont {Nandy}\ \emph {et~al.}(2012)\citenamefont {Nandy},
  \citenamefont {Santosh},\ and\ \citenamefont
  {Maiti}}]{santosh2012interaction}%
  \BibitemOpen
  \bibfield  {author} {\bibinfo {author} {\bibfnamefont {B.}~\bibnamefont
  {Nandy}}, \bibinfo {author} {\bibfnamefont {M.}~\bibnamefont {Santosh}}, \
  and\ \bibinfo {author} {\bibfnamefont {P.~K.}\ \bibnamefont {Maiti}},\
  }\href@noop {} {\bibfield  {journal} {\bibinfo  {journal} {J. Biosci.}\
  }\textbf {\bibinfo {volume} {37}},\ \bibinfo {pages} {457} (\bibinfo {year}
  {2012})}\BibitemShut {NoStop}%
\bibitem [{\citenamefont {Walt}(2013)}]{walt2013optical}%
  \BibitemOpen
  \bibfield  {author} {\bibinfo {author} {\bibfnamefont {D.~R.}\ \bibnamefont
  {Walt}},\ }\href@noop {} {\bibfield  {journal} {\bibinfo  {journal} {Anal.
  Chem.}\ }\textbf {\bibinfo {volume} {85}},\ \bibinfo {pages} {1258} (\bibinfo
  {year} {2013})}\BibitemShut {NoStop}%
\bibitem [{\citenamefont {Deniz}\ \emph {et~al.}(2008)\citenamefont {Deniz},
  \citenamefont {Mukhopadhyay},\ and\ \citenamefont {Lemke}}]{deniz2008single}%
  \BibitemOpen
  \bibfield  {author} {\bibinfo {author} {\bibfnamefont {A.~A.}\ \bibnamefont
  {Deniz}}, \bibinfo {author} {\bibfnamefont {S.}~\bibnamefont {Mukhopadhyay}},
  \ and\ \bibinfo {author} {\bibfnamefont {E.~A.}\ \bibnamefont {Lemke}},\
  }\href@noop {} {\bibfield  {journal} {\bibinfo  {journal} {J. R. Soc.
  Interface.}\ }\textbf {\bibinfo {volume} {5}},\ \bibinfo {pages} {15}
  (\bibinfo {year} {2008})}\BibitemShut {NoStop}%
\bibitem [{\citenamefont {Keyser}(2011)}]{keyser2011controlling}%
  \BibitemOpen
  \bibfield  {author} {\bibinfo {author} {\bibfnamefont {U.~F.}\ \bibnamefont
  {Keyser}},\ }\href@noop {} {\bibfield  {journal} {\bibinfo  {journal} {J. R.
  Soc. Interface.}\ }\textbf {\bibinfo {volume} {8}},\ \bibinfo {pages} {1369}
  (\bibinfo {year} {2011})}\BibitemShut {NoStop}%
\bibitem [{\citenamefont {Kim}\ \emph {et~al.}(2014)\citenamefont {Kim},
  \citenamefont {Kannam}, \citenamefont {Harrer}, \citenamefont {Downton},
  \citenamefont {Moore},\ and\ \citenamefont {Wagner}}]{kim2014@geometric}%
  \BibitemOpen
  \bibfield  {author} {\bibinfo {author} {\bibfnamefont {S.~C.}\ \bibnamefont
  {Kim}}, \bibinfo {author} {\bibfnamefont {S.~K.}\ \bibnamefont {Kannam}},
  \bibinfo {author} {\bibfnamefont {S.}~\bibnamefont {Harrer}}, \bibinfo
  {author} {\bibfnamefont {M.~T.}\ \bibnamefont {Downton}}, \bibinfo {author}
  {\bibfnamefont {S.}~\bibnamefont {Moore}}, \ and\ \bibinfo {author}
  {\bibfnamefont {J.~M.}\ \bibnamefont {Wagner}},\ }\href@noop {} {\bibfield
  {journal} {\bibinfo  {journal} {Phys. Rev. E}\ }\textbf {\bibinfo {volume}
  {89}},\ \bibinfo {pages} {042702} (\bibinfo {year} {2014})}\BibitemShut
  {NoStop}%
\bibitem [{\citenamefont {Carson}\ and\ \citenamefont
  {Wanunu}(2015)}]{carson2015challenges}%
  \BibitemOpen
  \bibfield  {author} {\bibinfo {author} {\bibfnamefont {S.}~\bibnamefont
  {Carson}}\ and\ \bibinfo {author} {\bibfnamefont {M.}~\bibnamefont
  {Wanunu}},\ }\href@noop {} {\bibfield  {journal} {\bibinfo  {journal}
  {Nanotech.}\ }\textbf {\bibinfo {volume} {26}},\ \bibinfo {pages} {074004}
  (\bibinfo {year} {2015})}\BibitemShut {NoStop}%
\bibitem [{\citenamefont {Venkatesan}\ and\ \citenamefont
  {Bashir}(2011)}]{venkatesan2011nanopore}%
  \BibitemOpen
  \bibfield  {author} {\bibinfo {author} {\bibfnamefont {B.~M.}\ \bibnamefont
  {Venkatesan}}\ and\ \bibinfo {author} {\bibfnamefont {R.}~\bibnamefont
  {Bashir}},\ }\href@noop {} {\bibfield  {journal} {\bibinfo  {journal} {Nat.
  Nanotech.}\ }\textbf {\bibinfo {volume} {6}},\ \bibinfo {pages} {615}
  (\bibinfo {year} {2011})}\BibitemShut {NoStop}%
\bibitem [{\citenamefont {Oukhaled}\ \emph {et~al.}(2012)\citenamefont
  {Oukhaled}, \citenamefont {Bacri}, \citenamefont {Pastoriza-Gallego},
  \citenamefont {Betton},\ and\ \citenamefont {Pelta}}]{oukhaled2012sensing}%
  \BibitemOpen
  \bibfield  {author} {\bibinfo {author} {\bibfnamefont {A.}~\bibnamefont
  {Oukhaled}}, \bibinfo {author} {\bibfnamefont {L.}~\bibnamefont {Bacri}},
  \bibinfo {author} {\bibfnamefont {M.}~\bibnamefont {Pastoriza-Gallego}},
  \bibinfo {author} {\bibfnamefont {J.-M.}\ \bibnamefont {Betton}}, \ and\
  \bibinfo {author} {\bibfnamefont {J.}~\bibnamefont {Pelta}},\ }\href@noop {}
  {\bibfield  {journal} {\bibinfo  {journal} {ACS Chem. Biol.}\ }\textbf
  {\bibinfo {volume} {7}},\ \bibinfo {pages} {1935} (\bibinfo {year}
  {2012})}\BibitemShut {NoStop}%
\bibitem [{\citenamefont {Ledden}\ \emph {et~al.}(2011)\citenamefont {Ledden},
  \citenamefont {Fologea}, \citenamefont {Talaga},\ and\ \citenamefont
  {Li}}]{ledden2011sensing}%
  \BibitemOpen
  \bibfield  {author} {\bibinfo {author} {\bibfnamefont {B.}~\bibnamefont
  {Ledden}}, \bibinfo {author} {\bibfnamefont {D.}~\bibnamefont {Fologea}},
  \bibinfo {author} {\bibfnamefont {D.~S.}\ \bibnamefont {Talaga}}, \ and\
  \bibinfo {author} {\bibfnamefont {J.}~\bibnamefont {Li}},\ }\enquote
  {\bibinfo {title} {Sensing single protein molecules with solid-state
  nanopores},}\ in\ \href@noop {} {\emph {\bibinfo {booktitle} {Nanopores:
  Sensing and Fundamental Biological Interactions}}}\ (\bibinfo  {publisher}
  {Springer US},\ \bibinfo {year} {2011})\ pp.\ \bibinfo {pages}
  {129--150}\BibitemShut {NoStop}%
\bibitem [{\citenamefont {Harrer}\ \emph {et~al.}(2015)\citenamefont {Harrer},
  \citenamefont {Kim}, \citenamefont {Schieber}, \citenamefont {Kannam},
  \citenamefont {Gunn}, \citenamefont {Moore}, \citenamefont {Scott},
  \citenamefont {Bathgate}, \citenamefont {Skafidas},\ and\ \citenamefont
  {Wagner}}]{harrer2015label}%
  \BibitemOpen
  \bibfield  {author} {\bibinfo {author} {\bibfnamefont {S.}~\bibnamefont
  {Harrer}}, \bibinfo {author} {\bibfnamefont {S.~C.}\ \bibnamefont {Kim}},
  \bibinfo {author} {\bibfnamefont {C.}~\bibnamefont {Schieber}}, \bibinfo
  {author} {\bibfnamefont {S.~K.}\ \bibnamefont {Kannam}}, \bibinfo {author}
  {\bibfnamefont {N.}~\bibnamefont {Gunn}}, \bibinfo {author} {\bibfnamefont
  {S.}~\bibnamefont {Moore}}, \bibinfo {author} {\bibfnamefont
  {S.}~\bibnamefont {Scott}}, \bibinfo {author} {\bibfnamefont
  {R.}~\bibnamefont {Bathgate}}, \bibinfo {author} {\bibfnamefont
  {S.}~\bibnamefont {Skafidas}}, \ and\ \bibinfo {author} {\bibfnamefont
  {J.~M.}\ \bibnamefont {Wagner}},\ }\href@noop {} {\bibfield  {journal}
  {\bibinfo  {journal} {Nanotech.}\ }\textbf {\bibinfo {volume} {26}},\
  \bibinfo {pages} {182502} (\bibinfo {year} {2015})}\BibitemShut {NoStop}%
\bibitem [{\citenamefont {Mohammad}\ and\ \citenamefont
  {Movileanu}(2012)}]{mohammad2012protein}%
  \BibitemOpen
  \bibfield  {author} {\bibinfo {author} {\bibfnamefont {M.~M.}\ \bibnamefont
  {Mohammad}}\ and\ \bibinfo {author} {\bibfnamefont {L.}~\bibnamefont
  {Movileanu}},\ }\enquote {\bibinfo {title} {Protein sensing with engineered
  protein nanopores},}\ in\ \href@noop {} {\emph {\bibinfo {booktitle}
  {Nanopore-Based Technology}}}\ (\bibinfo  {publisher} {Humana Press},\
  \bibinfo {year} {2012})\ pp.\ \bibinfo {pages} {21--37}\BibitemShut {NoStop}%
\bibitem [{\citenamefont {Larkin}\ \emph {et~al.}(2014)\citenamefont {Larkin},
  \citenamefont {Henley}, \citenamefont {Muthukumar}, \citenamefont
  {Rosenstein},\ and\ \citenamefont {Wanunu}}]{larkin2014high}%
  \BibitemOpen
  \bibfield  {author} {\bibinfo {author} {\bibfnamefont {J.}~\bibnamefont
  {Larkin}}, \bibinfo {author} {\bibfnamefont {R.~Y.}\ \bibnamefont {Henley}},
  \bibinfo {author} {\bibfnamefont {M.}~\bibnamefont {Muthukumar}}, \bibinfo
  {author} {\bibfnamefont {J.~K.}\ \bibnamefont {Rosenstein}}, \ and\ \bibinfo
  {author} {\bibfnamefont {M.}~\bibnamefont {Wanunu}},\ }\href@noop {}
  {\bibfield  {journal} {\bibinfo  {journal} {Biophys. J.}\ }\textbf {\bibinfo
  {volume} {106}},\ \bibinfo {pages} {696} (\bibinfo {year}
  {2014})}\BibitemShut {NoStop}%
\bibitem [{\citenamefont {Bonome}\ \emph {et~al.}(2015)\citenamefont {Bonome},
  \citenamefont {Lepore}, \citenamefont {Raimondo}, \citenamefont {Cecconi},
  \citenamefont {Tramontano},\ and\ \citenamefont
  {Chinappi}}]{bonome2015multistep}%
  \BibitemOpen
  \bibfield  {author} {\bibinfo {author} {\bibfnamefont {E.~L.}\ \bibnamefont
  {Bonome}}, \bibinfo {author} {\bibfnamefont {R.}~\bibnamefont {Lepore}},
  \bibinfo {author} {\bibfnamefont {D.}~\bibnamefont {Raimondo}}, \bibinfo
  {author} {\bibfnamefont {F.}~\bibnamefont {Cecconi}}, \bibinfo {author}
  {\bibfnamefont {A.}~\bibnamefont {Tramontano}}, \ and\ \bibinfo {author}
  {\bibfnamefont {M.}~\bibnamefont {Chinappi}},\ }\href@noop {} {\bibfield
  {journal} {\bibinfo  {journal} {J. Phys. Chem. B}\ }\textbf {\bibinfo
  {volume} {119}},\ \bibinfo {pages} {5815} (\bibinfo {year}
  {2015})}\BibitemShut {NoStop}%
\bibitem [{\citenamefont {Plesa}\ \emph {et~al.}(2013)\citenamefont {Plesa},
  \citenamefont {Kowalczyk}, \citenamefont {Zinsmeester}, \citenamefont
  {Grosberg}, \citenamefont {Rabin},\ and\ \citenamefont
  {Dekker}}]{plesa2013fast}%
  \BibitemOpen
  \bibfield  {author} {\bibinfo {author} {\bibfnamefont {C.}~\bibnamefont
  {Plesa}}, \bibinfo {author} {\bibfnamefont {S.~W.}\ \bibnamefont
  {Kowalczyk}}, \bibinfo {author} {\bibfnamefont {R.}~\bibnamefont
  {Zinsmeester}}, \bibinfo {author} {\bibfnamefont {A.~Y.}\ \bibnamefont
  {Grosberg}}, \bibinfo {author} {\bibfnamefont {Y.}~\bibnamefont {Rabin}}, \
  and\ \bibinfo {author} {\bibfnamefont {C.}~\bibnamefont {Dekker}},\
  }\href@noop {} {\bibfield  {journal} {\bibinfo  {journal} {Nano Lett.}\
  }\textbf {\bibinfo {volume} {13}},\ \bibinfo {pages} {658} (\bibinfo {year}
  {2013})}\BibitemShut {NoStop}%
\bibitem [{\citenamefont {Feng}\ \emph {et~al.}(2015)\citenamefont {Feng},
  \citenamefont {Zhang}, \citenamefont {Ying}, \citenamefont {Wang},\ and\
  \citenamefont {Du}}]{feng2015nanopore}%
  \BibitemOpen
  \bibfield  {author} {\bibinfo {author} {\bibfnamefont {Y.}~\bibnamefont
  {Feng}}, \bibinfo {author} {\bibfnamefont {Y.}~\bibnamefont {Zhang}},
  \bibinfo {author} {\bibfnamefont {C.}~\bibnamefont {Ying}}, \bibinfo {author}
  {\bibfnamefont {D.}~\bibnamefont {Wang}}, \ and\ \bibinfo {author}
  {\bibfnamefont {C.}~\bibnamefont {Du}},\ }\href@noop {} {\bibfield  {journal}
  {\bibinfo  {journal} {Genom. Proteom. Bioinf.}\ }\textbf {\bibinfo {volume}
  {13}},\ \bibinfo {pages} {4} (\bibinfo {year} {2015})}\BibitemShut {NoStop}%
\bibitem [{\citenamefont {Maitra}\ \emph {et~al.}(2012)\citenamefont {Maitra},
  \citenamefont {Kim},\ and\ \citenamefont {Dunbar}}]{maitra2012recent}%
  \BibitemOpen
  \bibfield  {author} {\bibinfo {author} {\bibfnamefont {R.~D.}\ \bibnamefont
  {Maitra}}, \bibinfo {author} {\bibfnamefont {J.}~\bibnamefont {Kim}}, \ and\
  \bibinfo {author} {\bibfnamefont {W.~B.}\ \bibnamefont {Dunbar}},\
  }\href@noop {} {\bibfield  {journal} {\bibinfo  {journal} {Electrophoresis}\
  }\textbf {\bibinfo {volume} {33}},\ \bibinfo {pages} {3418} (\bibinfo {year}
  {2012})}\BibitemShut {NoStop}%
\bibitem [{\citenamefont {Sung}\ and\ \citenamefont
  {Park}(1996)}]{sung1996polymer}%
  \BibitemOpen
  \bibfield  {author} {\bibinfo {author} {\bibfnamefont {W.}~\bibnamefont
  {Sung}}\ and\ \bibinfo {author} {\bibfnamefont {P.J.}~\bibnamefont {Park}},\
  }\href@noop {} {\bibfield  {journal} {\bibinfo  {journal} {Phys. Rev. Lett.}\
  }\textbf {\bibinfo {volume} {77}},\ \bibinfo {pages} {783} (\bibinfo {year}
  {1996})}\BibitemShut {NoStop}%
\bibitem [{\citenamefont {Muthukumar}(1999)}]{muthukumar1999polymer}%
  \BibitemOpen
  \bibfield  {author} {\bibinfo {author} {\bibfnamefont {M.}~\bibnamefont
  {Muthukumar}},\ }\href@noop {} {\bibfield  {journal} {\bibinfo  {journal} {J.
  Chem. Phys.}\ }\textbf {\bibinfo {volume} {111}},\ \bibinfo {pages} {10371}
  (\bibinfo {year} {1999})}\BibitemShut {NoStop}%
\bibitem [{\citenamefont {Kantor}\ and\ \citenamefont
  {Kardar}(2004)}]{kantor2004anomalous}%
  \BibitemOpen
  \bibfield  {author} {\bibinfo {author} {\bibfnamefont {Y.}~\bibnamefont
  {Kantor}}\ and\ \bibinfo {author} {\bibfnamefont {M.}~\bibnamefont
  {Kardar}},\ }\href@noop {} {\bibfield  {journal} {\bibinfo  {journal} {Phys.
  Rev. E}\ }\textbf {\bibinfo {volume} {69}},\ \bibinfo {pages} {021806}
  (\bibinfo {year} {2004})}\BibitemShut {NoStop}%
\bibitem [{\citenamefont {Storm}\ \emph {et~al.}(2005)\citenamefont {Storm},
  \citenamefont {Storm}, \citenamefont {Chen}, \citenamefont {Zandbergen},
  \citenamefont {Joanny},\ and\ \citenamefont {Dekker}}]{storm2005fast}%
  \BibitemOpen
  \bibfield  {author} {\bibinfo {author} {\bibfnamefont {A.~J.}\ \bibnamefont
  {Storm}}, \bibinfo {author} {\bibfnamefont {C.}~\bibnamefont {Storm}},
  \bibinfo {author} {\bibfnamefont {J.}~\bibnamefont {Chen}}, \bibinfo {author}
  {\bibfnamefont {H.}~\bibnamefont {Zandbergen}}, \bibinfo {author}
  {\bibfnamefont {J.-F.}\ \bibnamefont {Joanny}}, \ and\ \bibinfo {author}
  {\bibfnamefont {C.}~\bibnamefont {Dekker}},\ }\href@noop {} {\bibfield
  {journal} {\bibinfo  {journal} {Nano Lett.}\ }\textbf {\bibinfo {volume}
  {5}},\ \bibinfo {pages} {1193} (\bibinfo {year} {2005})}\BibitemShut
  {NoStop}%
\bibitem [{\citenamefont {Muthukumar}(2014)}]{muthukumar2014communication}%
  \BibitemOpen
  \bibfield  {author} {\bibinfo {author} {\bibfnamefont {M.}~\bibnamefont
  {Muthukumar}},\ }\href@noop {} {\bibfield  {journal} {\bibinfo  {journal} {J.
  Chem. Phys.}\ }\textbf {\bibinfo {volume} {141}},\ \bibinfo {pages} {081104}
  (\bibinfo {year} {2014})}\BibitemShut {NoStop}%
\bibitem [{\citenamefont {Fologea}\ \emph {et~al.}(2005)\citenamefont
  {Fologea}, \citenamefont {Uplinger}, \citenamefont {Thomas}, \citenamefont
  {McNabb},\ and\ \citenamefont {Li}}]{fologea2005slowing}%
  \BibitemOpen
  \bibfield  {author} {\bibinfo {author} {\bibfnamefont {D.}~\bibnamefont
  {Fologea}}, \bibinfo {author} {\bibfnamefont {J.}~\bibnamefont {Uplinger}},
  \bibinfo {author} {\bibfnamefont {B.}~\bibnamefont {Thomas}}, \bibinfo
  {author} {\bibfnamefont {D.~S.}\ \bibnamefont {McNabb}}, \ and\ \bibinfo
  {author} {\bibfnamefont {J.}~\bibnamefont {Li}},\ }\href@noop {} {\bibfield
  {journal} {\bibinfo  {journal} {Nano Lett.}\ }\textbf {\bibinfo {volume}
  {5}},\ \bibinfo {pages} {1734} (\bibinfo {year} {2005})}\BibitemShut
  {NoStop}%
\bibitem [{\citenamefont {Kannam}\ and\ \citenamefont
  {Downton}(2017)}]{kannam2017translational}%
  \BibitemOpen
  \bibfield  {author} {\bibinfo {author} {\bibfnamefont {S.~K.}\ \bibnamefont
  {Kannam}}\ and\ \bibinfo {author} {\bibfnamefont {M.~T.}\ \bibnamefont
  {Downton}},\ }\href@noop {} {\bibfield  {journal} {\bibinfo  {journal} {J.
  Chem. Phys.}\ }\textbf {\bibinfo {volume} {146}},\ \bibinfo {pages} {054108}
  (\bibinfo {year} {2017})}\BibitemShut {NoStop}%
\bibitem [{\citenamefont {Marrink}\ \emph {et~al.}(2004)\citenamefont
  {Marrink}, \citenamefont {de~Vries},\ and\ \citenamefont
  {Mark}}]{marrink2004coarse}%
  \BibitemOpen
  \bibfield  {author} {\bibinfo {author} {\bibfnamefont {S.~J.}\ \bibnamefont
  {Marrink}}, \bibinfo {author} {\bibfnamefont {A.~H.}\ \bibnamefont
  {de~Vries}}, \ and\ \bibinfo {author} {\bibfnamefont {A.~E.}\ \bibnamefont
  {Mark}},\ }\href@noop {} {\bibfield  {journal} {\bibinfo  {journal} {J. Phys.
  Chem. B}\ }\textbf {\bibinfo {volume} {108}},\ \bibinfo {pages} {750}
  (\bibinfo {year} {2004})}\BibitemShut {NoStop}%
\bibitem [{\citenamefont {Marrink}\ \emph {et~al.}(2007)\citenamefont
  {Marrink}, \citenamefont {Risselada}, \citenamefont {Yefimov}, \citenamefont
  {Tieleman},\ and\ \citenamefont {de~Vries}}]{marrink2007martini}%
  \BibitemOpen
  \bibfield  {author} {\bibinfo {author} {\bibfnamefont {S.~J.}\ \bibnamefont
  {Marrink}}, \bibinfo {author} {\bibfnamefont {H.~J.}\ \bibnamefont
  {Risselada}}, \bibinfo {author} {\bibfnamefont {S.}~\bibnamefont {Yefimov}},
  \bibinfo {author} {\bibfnamefont {D.~P.}\ \bibnamefont {Tieleman}}, \ and\
  \bibinfo {author} {\bibfnamefont {A.~H.}\ \bibnamefont {de~Vries}},\
  }\href@noop {} {\bibfield  {journal} {\bibinfo  {journal} {J. Phys. Chem. B}\
  }\textbf {\bibinfo {volume} {111}},\ \bibinfo {pages} {7812} (\bibinfo {year}
  {2007})}\BibitemShut {NoStop}%
\bibitem [{\citenamefont {Barrette-Ng}\ \emph {et~al.}(2013)\citenamefont
  {Barrette-Ng}, \citenamefont {Wu}, \citenamefont {Tjia}, \citenamefont
  {Wong},\ and\ \citenamefont {Ng}}]{Barrette-Ng2013the}%
  \BibitemOpen
  \bibfield  {author} {\bibinfo {author} {\bibfnamefont {I.~H.}\ \bibnamefont
  {Barrette-Ng}}, \bibinfo {author} {\bibfnamefont {S.-C.}\ \bibnamefont {Wu}},
  \bibinfo {author} {\bibfnamefont {W.-M.}\ \bibnamefont {Tjia}}, \bibinfo
  {author} {\bibfnamefont {S.-L.}\ \bibnamefont {Wong}}, \ and\ \bibinfo
  {author} {\bibfnamefont {K.~K.~S.}\ \bibnamefont {Ng}},\ }\href@noop {}
  {\bibfield  {journal} {\bibinfo  {journal} {Acta Crystallogra. Sect. D}\
  }\textbf {\bibinfo {volume} {69}},\ \bibinfo {pages} {879} (\bibinfo {year}
  {2013})}\BibitemShut {NoStop}%
\bibitem [{\citenamefont {D{\"u}nweg}\ and\ \citenamefont
  {Kremer}(1993)}]{dunweg1993molecular}%
  \BibitemOpen
  \bibfield  {author} {\bibinfo {author} {\bibfnamefont {B.}~\bibnamefont
  {D{\"u}nweg}}\ and\ \bibinfo {author} {\bibfnamefont {K.}~\bibnamefont
  {Kremer}},\ }\href@noop {} {\bibfield  {journal} {\bibinfo  {journal} {J.
  Chem. Phys.}\ }\textbf {\bibinfo {volume} {99}},\ \bibinfo {pages} {6983}
  (\bibinfo {year} {1993})}\BibitemShut {NoStop}%
\bibitem [{\citenamefont {Berendsen}\ \emph {et~al.}(1995)\citenamefont
  {Berendsen}, \citenamefont {van~der Spoel},\ and\ \citenamefont {van
  Drunen}}]{berendsen1995gromacs}%
  \BibitemOpen
  \bibfield  {author} {\bibinfo {author} {\bibfnamefont {H.~J.}\ \bibnamefont
  {Berendsen}}, \bibinfo {author} {\bibfnamefont {D.}~\bibnamefont {van~der
  Spoel}}, \ and\ \bibinfo {author} {\bibfnamefont {R.}~\bibnamefont {van
  Drunen}},\ }\href@noop {} {\bibfield  {journal} {\bibinfo  {journal} {Comput.
  Phys. Commun.}\ }\textbf {\bibinfo {volume} {91}},\ \bibinfo {pages} {43}
  (\bibinfo {year} {1995})}\BibitemShut {NoStop}%
\bibitem [{\citenamefont {Fologea}\ \emph {et~al.}(2007)\citenamefont
  {Fologea}, \citenamefont {Ledden}, \citenamefont {McNabb},\ and\
  \citenamefont {Li}}]{fologea2007electrical}%
  \BibitemOpen
  \bibfield  {author} {\bibinfo {author} {\bibfnamefont {D.}~\bibnamefont
  {Fologea}}, \bibinfo {author} {\bibfnamefont {B.}~\bibnamefont {Ledden}},
  \bibinfo {author} {\bibfnamefont {D.~S.}\ \bibnamefont {McNabb}}, \ and\
  \bibinfo {author} {\bibfnamefont {J.}~\bibnamefont {Li}},\ }\href@noop {}
  {\bibfield  {journal} {\bibinfo  {journal} {Appl. Phys. Lett.}\ }\textbf
  {\bibinfo {volume} {91}},\ \bibinfo {pages} {053901} (\bibinfo {year}
  {2007})}\BibitemShut {NoStop}%
\bibitem [{\citenamefont {Kannam}\ \emph {et~al.}(2014)\citenamefont {Kannam},
  \citenamefont {Kim}, \citenamefont {Rogers}, \citenamefont {Gunn},
  \citenamefont {Wagner}, \citenamefont {Harrer},\ and\ \citenamefont
  {Downtown}}]{kannam2014sensing}%
  \BibitemOpen
  \bibfield  {author} {\bibinfo {author} {\bibfnamefont {S.~K.}\ \bibnamefont
  {Kannam}}, \bibinfo {author} {\bibfnamefont {S.~C.}\ \bibnamefont {Kim}},
  \bibinfo {author} {\bibfnamefont {P.~R.}\ \bibnamefont {Rogers}}, \bibinfo
  {author} {\bibfnamefont {N.}~\bibnamefont {Gunn}}, \bibinfo {author}
  {\bibfnamefont {J.}~\bibnamefont {Wagner}}, \bibinfo {author} {\bibfnamefont
  {S.}~\bibnamefont {Harrer}}, \ and\ \bibinfo {author} {\bibfnamefont {M.~T.}\
  \bibnamefont {Downtown}},\ }\href@noop {} {\bibfield  {journal} {\bibinfo
  {journal} {Nanotech.}\ }\textbf {\bibinfo {volume} {25}},\ \bibinfo {pages}
  {155502} (\bibinfo {year} {2014})}\BibitemShut {NoStop}%
\bibitem [{\citenamefont {Talaga}\ and\ \citenamefont
  {Li}(2009)}]{talaga2009single}%
  \BibitemOpen
  \bibfield  {author} {\bibinfo {author} {\bibfnamefont {D.~S.}\ \bibnamefont
  {Talaga}}\ and\ \bibinfo {author} {\bibfnamefont {J.}~\bibnamefont {Li}},\
  }\href@noop {} {\bibfield  {journal} {\bibinfo  {journal} {J. Am. Chem.
  Soc.}\ }\textbf {\bibinfo {volume} {131}},\ \bibinfo {pages} {9287} (\bibinfo
  {year} {2009})}\BibitemShut {NoStop}%
\bibitem [{\citenamefont {Aksimentiev}(2010)}]{aksimentiev2009deciphering}%
  \BibitemOpen
  \bibfield  {author} {\bibinfo {author} {\bibfnamefont {A.}~\bibnamefont
  {Aksimentiev}},\ }\href@noop {} {\bibfield  {journal} {\bibinfo  {journal}
  {Nanoscale}\ }\textbf {\bibinfo {volume} {2}},\ \bibinfo {pages} {468}
  (\bibinfo {year} {2010})}\BibitemShut {NoStop}%
\bibitem [{\citenamefont {Gianti}\ \emph {et~al.}(2016)\citenamefont {Gianti},
  \citenamefont {Delemotte}, \citenamefont {Klein},\ and\ \citenamefont
  {Carnevale}}]{eleonora2016role}%
  \BibitemOpen
  \bibfield  {author} {\bibinfo {author} {\bibfnamefont {E.}~\bibnamefont
  {Gianti}}, \bibinfo {author} {\bibfnamefont {L.}~\bibnamefont {Delemotte}},
  \bibinfo {author} {\bibfnamefont {M.~L.}\ \bibnamefont {Klein}}, \ and\
  \bibinfo {author} {\bibfnamefont {V.}~\bibnamefont {Carnevale}},\ }\href@noop
  {} {\bibfield  {journal} {\bibinfo  {journal} {Proc. Natl. Acad. Sci. USA.}\
  }\textbf {\bibinfo {volume} {113}},\ \bibinfo {pages} {8359} (\bibinfo {year}
  {2016})}\BibitemShut {NoStop}%
\bibitem [{\citenamefont {van Keulen}\ \emph {et~al.}(2017)\citenamefont {van
  Keulen}, \citenamefont {Gianti}, \citenamefont {Carnevale}, \citenamefont
  {Klein}, \citenamefont {Rothlisberger},\ and\ \citenamefont
  {Delemotte}}]{eleonora2017proton}%
  \BibitemOpen
  \bibfield  {author} {\bibinfo {author} {\bibfnamefont {S.~C.}\ \bibnamefont
  {van Keulen}}, \bibinfo {author} {\bibfnamefont {E.}~\bibnamefont {Gianti}},
  \bibinfo {author} {\bibfnamefont {V.}~\bibnamefont {Carnevale}}, \bibinfo
  {author} {\bibfnamefont {M.~L.}\ \bibnamefont {Klein}}, \bibinfo {author}
  {\bibfnamefont {U.}~\bibnamefont {Rothlisberger}}, \ and\ \bibinfo {author}
  {\bibfnamefont {L.}~\bibnamefont {Delemotte}},\ }\href@noop {} {\bibfield
  {journal} {\bibinfo  {journal} {J. Phys. Chem. B}\ }\textbf {\bibinfo
  {volume} {121}},\ \bibinfo {pages} {3340} (\bibinfo {year}
  {2017})}\BibitemShut {NoStop}%
\bibitem [{\citenamefont {Rodriguez}\ and\ \citenamefont
  {Li}(1999)}]{rodriguez1991surface}%
  \BibitemOpen
  \bibfield  {author} {\bibinfo {author} {\bibfnamefont {I.}~\bibnamefont
  {Rodriguez}}\ and\ \bibinfo {author} {\bibfnamefont {S.~F.~Y.}\ \bibnamefont
  {Li}},\ }\href@noop {} {\bibfield  {journal} {\bibinfo  {journal} {Anal.
  Chim. Acta}\ }\textbf {\bibinfo {volume} {383}},\ \bibinfo {pages} {1}
  (\bibinfo {year} {1999})}\BibitemShut {NoStop}%
\bibitem [{\citenamefont {Durand}\ and\ \citenamefont
  {Philippe}(2009)}]{durand2008label}%
  \BibitemOpen
  \bibfield  {author} {\bibinfo {author} {\bibfnamefont {N.~F.~Y.}\
  \bibnamefont {Durand}}\ and\ \bibinfo {author} {\bibfnamefont
  {R.}~\bibnamefont {Philippe}},\ }\href@noop {} {\bibfield  {journal}
  {\bibinfo  {journal} {Lab on a Chip}\ }\textbf {\bibinfo {volume} {9}},\
  \bibinfo {pages} {319} (\bibinfo {year} {2009})}\BibitemShut {NoStop}%
\bibitem [{\citenamefont {Carr}\ \emph {et~al.}(2011)\citenamefont {Carr},
  \citenamefont {Comer}, \citenamefont {Ginsberg},\ and\ \citenamefont
  {Aksimentiev}}]{carr2011modelling}%
  \BibitemOpen
  \bibfield  {author} {\bibinfo {author} {\bibfnamefont {R.}~\bibnamefont
  {Carr}}, \bibinfo {author} {\bibfnamefont {J.}~\bibnamefont {Comer}},
  \bibinfo {author} {\bibfnamefont {M.~D.}\ \bibnamefont {Ginsberg}}, \ and\
  \bibinfo {author} {\bibfnamefont {A.}~\bibnamefont {Aksimentiev}},\
  }\href@noop {} {\bibfield  {journal} {\bibinfo  {journal} {IEEE Trans.
  Nanotech.}\ }\textbf {\bibinfo {volume} {10}},\ \bibinfo {pages} {75}
  (\bibinfo {year} {2011})}\BibitemShut {NoStop}%
\bibitem [{\citenamefont {Student}(1908)}]{student1908probable}%
  \BibitemOpen
  \bibfield  {author} {\bibinfo {author} {\bibnamefont {Student}},\ }\href@noop
  {} {\bibfield  {journal} {\bibinfo  {journal} {Biometrika}\ ,\ \bibinfo
  {pages} {1}} (\bibinfo {year} {1908})}\BibitemShut {NoStop}%
\bibitem [{\citenamefont {Ikonen}\ \emph {et~al.}(2013)\citenamefont {Ikonen},
  \citenamefont {Bhattacharya}, \citenamefont {Ala-Nissila},\ and\
  \citenamefont {Sung}}]{ikonen2013influence}%
  \BibitemOpen
  \bibfield  {author} {\bibinfo {author} {\bibfnamefont {T.}~\bibnamefont
  {Ikonen}}, \bibinfo {author} {\bibfnamefont {A.}~\bibnamefont
  {Bhattacharya}}, \bibinfo {author} {\bibfnamefont {T.}~\bibnamefont
  {Ala-Nissila}}, \ and\ \bibinfo {author} {\bibfnamefont {W.}~\bibnamefont
  {Sung}},\ }\href@noop {} {\bibfield  {journal} {\bibinfo  {journal} {EPL
  (Europhys. Lett.)}\ }\textbf {\bibinfo {volume} {103}},\ \bibinfo {pages}
  {38001} (\bibinfo {year} {2013})}\BibitemShut {NoStop}%
\bibitem [{\citenamefont {Stellwagen}\ \emph {et~al.}(2001)\citenamefont
  {Stellwagen}, \citenamefont {Bossi}, \citenamefont {Gelfi},\ and\
  \citenamefont {Righetti}}]{stellwagen2001orientation}%
  \BibitemOpen
  \bibfield  {author} {\bibinfo {author} {\bibfnamefont {N.~C.}\ \bibnamefont
  {Stellwagen}}, \bibinfo {author} {\bibfnamefont {A.}~\bibnamefont {Bossi}},
  \bibinfo {author} {\bibfnamefont {C.}~\bibnamefont {Gelfi}}, \ and\ \bibinfo
  {author} {\bibfnamefont {P.~G.}\ \bibnamefont {Righetti}},\ }\href@noop {}
  {\bibfield  {journal} {\bibinfo  {journal} {Electrophoresis}\ }\textbf
  {\bibinfo {volume} {22}},\ \bibinfo {pages} {4311} (\bibinfo {year}
  {2001})}\BibitemShut {NoStop}%
\bibitem [{\citenamefont {Chen}\ \emph {et~al.}(2004)\citenamefont {Chen},
  \citenamefont {Gu}, \citenamefont {Brandin}, \citenamefont {Kim},
  \citenamefont {Wang},\ and\ \citenamefont {Branton}}]{peng2004probing}%
  \BibitemOpen
  \bibfield  {author} {\bibinfo {author} {\bibfnamefont {P.}~\bibnamefont
  {Chen}}, \bibinfo {author} {\bibfnamefont {J.}~\bibnamefont {Gu}}, \bibinfo
  {author} {\bibfnamefont {E.}~\bibnamefont {Brandin}}, \bibinfo {author}
  {\bibfnamefont {Y.-R.}\ \bibnamefont {Kim}}, \bibinfo {author} {\bibfnamefont
  {Q.}~\bibnamefont {Wang}}, \ and\ \bibinfo {author} {\bibfnamefont
  {D.}~\bibnamefont {Branton}},\ }\href@noop {} {\bibfield  {journal} {\bibinfo
   {journal} {Nano Lett.}\ }\textbf {\bibinfo {volume} {4}},\ \bibinfo {pages}
  {2293} (\bibinfo {year} {2004})}\BibitemShut {NoStop}%
\bibitem [{\citenamefont {Brown}\ and\ \citenamefont
  {Rymden}(1988)}]{brown1988comparison}%
  \BibitemOpen
  \bibfield  {author} {\bibinfo {author} {\bibfnamefont {W.}~\bibnamefont
  {Brown}}\ and\ \bibinfo {author} {\bibfnamefont {R.}~\bibnamefont {Rymden}},\
  }\href@noop {} {\bibfield  {journal} {\bibinfo  {journal} {Macromol.}\
  }\textbf {\bibinfo {volume} {21}},\ \bibinfo {pages} {840} (\bibinfo {year}
  {1988})}\BibitemShut {NoStop}%
\bibitem [{\citenamefont {Pryamitsyn}\ and\ \citenamefont
  {Ganesan}(2016)}]{pryamitsyn2016noncontinuum}%
  \BibitemOpen
  \bibfield  {author} {\bibinfo {author} {\bibfnamefont {V.}~\bibnamefont
  {Pryamitsyn}}\ and\ \bibinfo {author} {\bibfnamefont {V.}~\bibnamefont
  {Ganesan}},\ }\href@noop {} {\bibfield  {journal} {\bibinfo  {journal} {J.
  Polym. Sci. : Polym. Phys. Ed.}\ }\textbf {\bibinfo {volume} {54}},\ \bibinfo
  {pages} {2145} (\bibinfo {year} {2016})}\BibitemShut {NoStop}%
\bibitem [{\citenamefont {Nkodo}\ \emph {et~al.}(2001)\citenamefont {Nkodo},
  \citenamefont {Garnier}, \citenamefont {Tinland}, \citenamefont {Ren},
  \citenamefont {Desruisseaux}, \citenamefont {McCormick}, \citenamefont
  {Drouin},\ and\ \citenamefont {Slater}}]{nkodo2001diffusion}%
  \BibitemOpen
  \bibfield  {author} {\bibinfo {author} {\bibfnamefont {A.~E.}\ \bibnamefont
  {Nkodo}}, \bibinfo {author} {\bibfnamefont {J.~M.}\ \bibnamefont {Garnier}},
  \bibinfo {author} {\bibfnamefont {B.}~\bibnamefont {Tinland}}, \bibinfo
  {author} {\bibfnamefont {H.}~\bibnamefont {Ren}}, \bibinfo {author}
  {\bibfnamefont {C.}~\bibnamefont {Desruisseaux}}, \bibinfo {author}
  {\bibfnamefont {L.~C.}\ \bibnamefont {McCormick}}, \bibinfo {author}
  {\bibfnamefont {G.}~\bibnamefont {Drouin}}, \ and\ \bibinfo {author}
  {\bibfnamefont {G.~W.}\ \bibnamefont {Slater}},\ }\href@noop {} {\bibfield
  {journal} {\bibinfo  {journal} {Electrophoresis}\ }\textbf {\bibinfo {volume}
  {22}},\ \bibinfo {pages} {2424} (\bibinfo {year} {2001})}\BibitemShut
  {NoStop}%
\bibitem [{\citenamefont {Carson}\ \emph {et~al.}(2014)\citenamefont {Carson},
  \citenamefont {Wilson}, \citenamefont {Aksimentiev},\ and\ \citenamefont
  {Wanunu}}]{carson2014smooth}%
  \BibitemOpen
  \bibfield  {author} {\bibinfo {author} {\bibfnamefont {S.}~\bibnamefont
  {Carson}}, \bibinfo {author} {\bibfnamefont {J.}~\bibnamefont {Wilson}},
  \bibinfo {author} {\bibfnamefont {A.}~\bibnamefont {Aksimentiev}}, \ and\
  \bibinfo {author} {\bibfnamefont {M.}~\bibnamefont {Wanunu}},\ }\href@noop {}
  {\bibfield  {journal} {\bibinfo  {journal} {Biophys. J.}\ }\textbf {\bibinfo
  {volume} {107}},\ \bibinfo {pages} {2381} (\bibinfo {year}
  {2014})}\BibitemShut {NoStop}%
\bibitem [{\citenamefont {Lubensky}\ and\ \citenamefont
  {Nelson}(1999)}]{lubensky1999driven}%
  \BibitemOpen
  \bibfield  {author} {\bibinfo {author} {\bibfnamefont {D.~K.}\ \bibnamefont
  {Lubensky}}\ and\ \bibinfo {author} {\bibfnamefont {D.~R.}\ \bibnamefont
  {Nelson}},\ }\href@noop {} {\bibfield  {journal} {\bibinfo  {journal}
  {Biophys. J.}\ }\textbf {\bibinfo {volume} {77}},\ \bibinfo {pages} {1824}
  (\bibinfo {year} {1999})}\BibitemShut {NoStop}%
\bibitem [{\citenamefont {Berezhkovskii}\ and\ \citenamefont
  {Gopich}(2003)}]{berezhkovskii2003translocation}%
  \BibitemOpen
  \bibfield  {author} {\bibinfo {author} {\bibfnamefont {A.}~\bibnamefont
  {Berezhkovskii}}\ and\ \bibinfo {author} {\bibfnamefont {I.}~\bibnamefont
  {Gopich}},\ }\href@noop {} {\bibfield  {journal} {\bibinfo  {journal}
  {Biophys. J.}\ }\textbf {\bibinfo {volume} {84}},\ \bibinfo {pages} {787}
  (\bibinfo {year} {2003})}\BibitemShut {NoStop}%
\bibitem [{\citenamefont {Ling}\ and\ \citenamefont {S}(2013)}]{ling2013on}%
  \BibitemOpen
  \bibfield  {author} {\bibinfo {author} {\bibfnamefont {D.~Y.}\ \bibnamefont
  {Ling}}\ and\ \bibinfo {author} {\bibfnamefont {L.~X.}\ \bibnamefont {S}},\
  }\href@noop {} {\bibfield  {journal} {\bibinfo  {journal} {J. Phys.: Cond.
  Matt.}\ }\textbf {\bibinfo {volume} {25}},\ \bibinfo {pages} {375102}
  (\bibinfo {year} {2013})}\BibitemShut {NoStop}%
\bibitem [{\citenamefont {Talaga}\ and\ \citenamefont
  {Li}(2013)}]{talaga2013correction}%
  \BibitemOpen
  \bibfield  {author} {\bibinfo {author} {\bibfnamefont {D.~S.}\ \bibnamefont
  {Talaga}}\ and\ \bibinfo {author} {\bibfnamefont {J.}~\bibnamefont {Li}},\
  }\href@noop {} {\bibfield  {journal} {\bibinfo  {journal} {J. Chem. Soc.}\
  }\textbf {\bibinfo {volume} {135}},\ \bibinfo {pages} {13220} (\bibinfo
  {year} {2013})}\BibitemShut {NoStop}%
\bibitem [{\citenamefont {Wanunu}\ \emph {et~al.}(2008)\citenamefont {Wanunu},
  \citenamefont {Sutin}, \citenamefont {McNally}, \citenamefont {Chow},\ and\
  \citenamefont {Meller}}]{wanunu2008dna}%
  \BibitemOpen
  \bibfield  {author} {\bibinfo {author} {\bibfnamefont {M.}~\bibnamefont
  {Wanunu}}, \bibinfo {author} {\bibfnamefont {J.}~\bibnamefont {Sutin}},
  \bibinfo {author} {\bibfnamefont {B.}~\bibnamefont {McNally}}, \bibinfo
  {author} {\bibfnamefont {A.}~\bibnamefont {Chow}}, \ and\ \bibinfo {author}
  {\bibfnamefont {A.}~\bibnamefont {Meller}},\ }\href@noop {} {\bibfield
  {journal} {\bibinfo  {journal} {Biophys. J.}\ }\textbf {\bibinfo {volume}
  {95}},\ \bibinfo {pages} {4716} (\bibinfo {year} {2008})}\BibitemShut
  {NoStop}%
\bibitem [{\citenamefont {Menais}\ \emph {et~al.}(2016)\citenamefont {Menais},
  \citenamefont {Mossa},\ and\ \citenamefont {Buhot}}]{menais2016polymer}%
  \BibitemOpen
  \bibfield  {author} {\bibinfo {author} {\bibfnamefont {T.}~\bibnamefont
  {Menais}}, \bibinfo {author} {\bibfnamefont {S.}~\bibnamefont {Mossa}}, \
  and\ \bibinfo {author} {\bibfnamefont {A.}~\bibnamefont {Buhot}},\
  }\href@noop {} {\bibfield  {journal} {\bibinfo  {journal} {Sci. Rep.}\
  }\textbf {\bibinfo {volume} {6}},\ \bibinfo {pages} {38558} (\bibinfo {year}
  {2016})}\BibitemShut {NoStop}%
\bibitem [{\citenamefont {Menais}(2017)}]{menais2017polymer}%
  \BibitemOpen
  \bibfield  {author} {\bibinfo {author} {\bibfnamefont {T.}~\bibnamefont
  {Menais}},\ }\href@noop {} {\bibfield  {journal} {\bibinfo  {journal}
  {Cond.Matt.arXiv}\ }\textbf {\bibinfo {volume} {1711.10832}} (\bibinfo {year}
  {2017})}\BibitemShut {NoStop}%
\end{thebibliography}

%
\end{document}